\documentclass[12pt]{article}

\usepackage[dvips]{graphics}

\begin{document}

\title{\bf On search for new Higgs physics in CDF at the Tevatron}
\author{
G.A. Kozlov\\
\em Bogoliubov Laboratory of Theoretical Physics,\\
\em Joint Institute for Nuclear Research,\\
\em 141980 Dubna, Moscow Region, Russia\\
\em e-mail: kozlov@thsun1.jinr.ru\\
T. Morii\\
\em Div. of Sciences for Natural Environment,\\
\em Faculty of Human Development,\\
\em Kobe University, Kobe, Japan\\
\em e-mail: morii@kobe-u.ac.jp }

\date{}
\maketitle
\begin{abstract}

{\small We discuss the Higgs boson mass sum rules in the Minimal
Supersymmetric Standard Model in order to estimate the upper
limits on the masses of stop quarks as well as the lower bounds
on the masses of the scalar Higgs boson state.
The bounds on the scale of quark-lepton
compositeness derived from the CDF Collaboration (Fermilab Tevatron)
 data and
applied to new extra gauge boson search is taken into account.
These extra gauge bosons are considered in the framework of
the extended $SU(2)_{h}\times SU(2)_{l}$ model. In addition,
we discuss the physics of rare decays of the MSSM Higgs bosons in both
CP-even and CP-odd sectors and also some extra gauge bosons.}

\end{abstract}

\section{Introduction}

There are still some serious ingredients in the fundamental
interactions of elementary particles that have not been
experimentally verified.
The Higgs particle(s) and new heavy neutral gauge bosons
$G^{\prime}\subset Z^{\prime}$, $W^{{\pm}^{\prime}}$,
$ Z^{\prime\prime}$, $W^{{\pm}^{\prime\prime}}$, ... have not
yet been established and physics of those particles still
remain elusive. In recent years, the amount of works for
searching for the Higgs and extra gauge bosons have considerably
increased and this research subject is now one of the most exciting
topics in searching physics beyond the standard model (SM).
On one hand, recent LEP 2 experiments determined a lower bound of
the Higgs boson mass to be approximately 114.1 GeV [1].
Furthermore, it should be pointed out that the
Tevatron data [2] for searching for the low energy effects of
quark-lepton contact interactions on dilepton production taken
at $\sqrt s$=1.8 TeV are translated into lower bounds on the masses of
extra neutral gauge bosons $Z^{\prime}$. On the other hand, recent
theoretical progress on study of strong and electroweak interactions and
the fundamental concept on natural extensions of the SM,
have turned the physics of interplay between the
known matter fields and exciting new phenomena related to
new matter fields via new interactions.

One of the most challenging current topics beyond the SM is to study
physical
implication of  a set of Higgs particles and extra
gauge bosons predicted by the Minimal
Supersymmetric Standard Model (MSSM) in the Tevatron $\bar{p}p$
collider experiment at $\sqrt s$=2 TeV.
The production of Higgs bosons
or their decays (via the heavy quark/lepton interactions) are
promising and useful processes which would be
searched for in the forthcoming high energy experiments.
With the advent of the Run II at the Tevatron, it is expected to be
clarified that the intermediate heavy quark loop can be a
sizeable source of Higgs bosons in both CP-even (light $h$ and
heavy $H$) scalar and CP-odd pseudoscalar ($A$) sectors.
As for extra neutral gauge bosons $G^{\prime}$, existence of those
neutral gauge bosons are
required by the models addressing the physics beyond the SM, such
as Supersymmetry (SUSY), Grand Unification Theory, superstrings and
so on. The prediction for the masses $M_{G^{\prime}}$ of
 $G^{\prime}$-bosons is rather uncertain, though
these bosons  could be heavy enough with the
masses $M_{G^{\prime}}\gg {\cal O}(m_{Z})$
($m_{Z}$ is the mass of the $Z$-boson). The simplest version
of existing $G^{\prime}$-bosons
is based on the model with extension of the standard $SU(3)\times
SU(2)\times U(1)$ gauge group by an extra $U(1)$ factor [3].
There are also many other extended models of the SM,
such as the ones in which the
$SU(2)$ gauge group is extended to $SU(2)\times SU(2)$ [4-7] and
to $SU(2)\times SU(2)\times U(1)$ [8]. In those
models, the massive SU(2) extra gauge
bosons (corresponding to the broken generators) could couple to
fermions in different generations with different strength and thus could give
the answer to the question of why the top quark is so
heavy, since they might single out the fermions of the third
generation.
The precision measurements of electroweak parameters narrowed the
allowed region of extra gauge boson masses keeping Higgs boson masses
 to be consistent with
radiative corrections including the supersymmetric ones.


One of the goals of the present work is to examine the potential
of the Run II experiment at the Tevatron in searching for $h$-, $H$-
and $A$- Higgs bosons, new extra gauge bosons, and to give the
estimation of the upper limits on the masses of stop quarks as
well as the lower bounds on the masses of the scalar Higgs bosons
both in the light and heavy mass sectors.

We first show how the existing Tevatron
bounds on the scale of quark-lepton compositeness [9] can be
adopted to provide an upper limit of the quantity
$m_{{\tilde t}_{1}}\cdot m_{{\tilde t}_{2}}$, i.e.
 the product of
masses of stop eigenstates ${\tilde t}_{1}$ and  ${\tilde
t}_{2}$. We also discuss how the lower bounds of
the scalar Higgs boson masses can be obtained
 from the forthcoming Tevatron data.
Furthermore, we
emphasize that the forthcoming experiments for discovering the
supersymmetry in both Higgs and quark sectors could lead to
estimations of the masses of neutral and charged extra gauge
bosons $Z^{\prime}$ and   $W^{{\pm}^{\prime}}$, respectively.
An example-the models in which the precision electroweak data allow the
extra gauge bosons to be of the order ${\cal O}$(0.5~ TeV) are, e.g.,
the noncommuting extended technicolor models [5].
The $Z^{\prime}$ and   $W^{{\pm}^{\prime}}$ bosons with such
masses are of interest, since they are within the kinematic
reach of the Tevatron's Run II experiments.

Then we also study the processes like
\begin{eqnarray}
\label{e1}
\bar{p}p\rightarrow gg\rightarrow (h/H) X, A X
\end{eqnarray}
keeping in mind that the main channels in the MSSM are the scalar
ones , i.e., $\bar{l}l$, $\bar{Q}Q$, $\gamma\gamma h$, $ggh$
($\gamma$, $g$, $l$ and $Q$ mean a photon, a gluon, a lepton
and a heavy quark, respectively) since the pseudoscalar mode is
largely suppressed in the wide range of the MSSM parameters.
Since in two photons ($\gamma\gamma$) or two gluons  ($gg$) decays the
invariant mass of $\gamma\gamma$ or $gg$ system would be
identical to the mass of the decaying boson, a promising
way to detect $h$- and $H$-bosons at the Tevatron is to
search for the decays $h$, $H\rightarrow \gamma\gamma$
and  $h$, $H\rightarrow gg$ as well.
It is known that at high energy $\bar{p}p$ (or $pp$) collisions
a contribution of gluonic interactions become large due to increase
of gluon densities in the proton.
The Tevatron Run II experiment could observe
 the lightest Higgs boson production via a two gluon
fusion $gg\rightarrow h$ with the cross-section $\sigma$ of an
order $\sigma\sim {\cal O}$(1.0 pb) at $m_{h}\sim$ 110 GeV [10].
Raise of the $h$-boson mass up to 180 GeV would lead to
decreasing of $\sigma$ up to the order ${\cal O}$(0.2 pb).
For the $A$-Higgs boson production the
following important features
are remarkable :\\
(i) the detection efficiency of the signal events has high
accuracy because the decay of $A$ Higgs-bosons can be precisely modeled
in the kinematic region for various decay channels
, e.g., $A\rightarrow \bar{l}l, \gamma\gamma h, ggh$
(here, the leptons $l$ run over electrons $e$, muons $\mu$ and
$\tau$-leptons);\\
(ii) the mass $M_{A}$ of the $A$ Higgs-boson  can be reconstructed from
its final state to test the mass relation
in the MSSM mass sum rule [11] at the tree-level
\begin{eqnarray}
\label{e2}
m_{h}^2+M_{H}^2=M_{A}^2+m_{Z}^2
\end{eqnarray}
with its deviation due to radiative corrections
($m_{h}$ and $M_{H}$ are the masses of the
 CP-even $h$ and $H$ Higgs bosons, respectively).

The outline of the article is as follows. In Section 2, we
define the model. Estimation of the upper limits of
the masses of stop quarks as well as the lower bounds on the
masses of the scalar Higgs bosons will be discussed in Sec. 3.
Section 4 is devoted to study on rare decays of
$h$-, $H$- and $A$- Higgs bosons in the MSSM. Section 5 focuses
on the $h$ Higgs boson production in the decay of an extra
gauge boson $Z_{2}$. Finally, in Section 6, we give our conclusions.

\section{The effective model. }

In the model of extended weak interactions governed by a pair of
$SU(2)$ gauge groups $SU(2)_{h}\times SU(2)_{l}$ for heavy
(third generation) and light fermions (labels $h$ and $l$ mean
heavy and light, respectively) the gauge boson eigenstates are
given by [12]
\begin{eqnarray}
\label{e3}
A^{\mu}=\sin\theta\,(\cos\phi\,W_{{3_{h}}}^{\mu}+\sin\phi\,W_{{3_{l}}}^{\mu})+
\cos\theta\,X^{\mu}
\end{eqnarray}
for a photon and
\begin{eqnarray}
\label{e4}
Z_{1}^{\mu}=\cos\theta\,(\cos\phi\,W_{{3_{h}}}^{\mu}+\sin\phi\,W_{{3_{l}}}^{\mu})-
\sin\theta\,X^{\mu}\ ,
\end{eqnarray}
\begin{eqnarray}
\label{e5}
Z_{2}^{\mu}=-\sin\phi\,W_{{3_{h}}}^{\mu}+\cos\phi\,W_{{3_{l}}}^{\mu}
\end{eqnarray}
for neutral gauge bosons $Z_{1}$, $Z_{2}$, respectively, which define
neutral mass eigenstates $Z$ and $Z^{\prime}$ at the leading order of a
free parameter $x$ [9]
\begin{eqnarray}
\label{e6}
{Z\choose Z^{\prime}}\simeq
\left(\matrix{1&\frac{-\cos^3\phi\,\sin\phi}{x\,\cos\theta}\cr
\frac{\cos^3\phi\,\sin\phi}{x\,\cos\theta}&1\cr}\right)\cdot
{Z_{1}\choose Z_{2}}\ ,
\end{eqnarray}
where $\theta$ is the usual weak
mixing angle and  $\phi$ is an additional mixing angle due to
the existence of $SU(2)_{h}\times SU(2)_{l}$.

The parameter $x$ in (\ref{e6}) is defined as the ratio
 $x=u^2/v^2$, where $u$ is the
energy scale at which the extended weak gauge group $SU(2)_{h}\times SU(2)_{l}$
is broken to its diagonal subgroup $SU(2)_{L}$, while $v\simeq$
246 GeV is the vacuum expectation value of the (composite)
scalar field responsible for the symmetry breaking $
SU(2)_{L}\times U(1)_{Y}\rightarrow U(1)_{em}$ in the model of
extended weak interactions. The generator of the
$U(1)_{em}$ group is the usual electric charge operator
$Q=T_{3_{h}}+T_{3_{l}}+Y$. At large values of $\sin\phi$, the
$Z_{2}$-boson has an enhanced coupling to the third generation
fermions through the covariant derivative
\begin{eqnarray}
\label{e7}
D^{\mu}=\partial^{\mu}-i\,\frac{g}{\cos\theta}\,Z_{1}^{\mu}\,\left(T_{3_{h}}+
T_{3_{l}}-\sin^{2}\theta\cdot Q\right)\cr
-i\,g\,Z_{2}^{\mu}\,\left(-\frac{\sin\phi}{\cos\phi}\,
T_{3_{h}}+\frac{\cos\phi}{\sin\phi}\,T_{3_{l}}\right).
\end{eqnarray}
 The Lagrangian density for an effective quark-lepton
 contact interaction looks like
\begin{eqnarray}
\label{e8}
{\cal L}\supset\frac{1}{\Lambda_{LL}^2}\left [g_{0}^2
(\bar{E}_{L}\gamma_{\mu} E_{L})(\bar{Q}_{L}\gamma^{\mu}
Q_{L})+g_{1}^2 (\bar{E}_{L}\gamma_{\mu}\tau_{a} E_{L})(\bar{Q}_{L}\gamma^{\mu}
\tau_{a} Q_{L})\right ]\cr
+\frac{g_{e}^2}{\Lambda_{LR}^2}
(\bar{e}_{R}\gamma_{\mu} e_{R})(\bar{Q}_{L}\gamma^{\mu}Q_{L})\cr
+\left [\frac{1}{\Lambda_{LR}^2}(\bar{E}_{L}\gamma_{\mu}E_{L})
+\frac{1}{\Lambda_{RR}^2}(\bar{e}_{R}\gamma_{\mu}
e_{R}) \right ]\sum_{q:u,d} g_{q}^2 (\bar{q}_{R}\gamma^{\mu}q_{R})\
,
\end{eqnarray}
where $E_{L}=(\nu_{e},e)_{L}, Q_{L}=(u,d)_{L}$; $g_{i}$ are the
effective couplings and $\Lambda_{ij}$ are the scales of new
physics. The aim of the CDF collaboration analysis [2] was to
search for the deviation of the SM prediction in the dilepton
production spectrum. If no such deviations have been found, the
lower bound of the $\Lambda$-scale can be obtained.
The embedding of the extra gauge bosons in the model
beyond the SM gives rise
to quark-lepton contact interactions in accordance to the
following part of the Lagrangian density (see [9])
\begin{eqnarray}
\label{e9}
{\cal L}\supset
-\frac{g^2}{M_{Z^{\prime}}^2}\left (\frac{\cot\phi}{2}\right )^2
\left (\sum_{l:e,\mu} \bar{l}_{L}\gamma_{\mu}l_{L}\right )
\left (\sum_{q:u,d,s,c} \bar{q}_{L}\gamma^{\mu}q_{L}\right )\
,
\end{eqnarray}
where $g=e/\sin\theta$.
We suppose that the couplings in the first two generations are
same in strength.

\section{Mass bound on stop quarks and some Higgs boson mass estimations}

 In the MSSM, the mass sum rule (\ref{e2}) at the tree-level
is transformed into the following form because of the loop corrections [13]
\begin{eqnarray}
\label{e10}
M_{Z^{\prime}}=\frac{m_{h}^2-M_{A}^2+\delta_{ZZ^{\prime}}
-\Delta}{M_{Z^{\prime}}
+M_{H}}+M_{H}\ ,
\end{eqnarray}
where $M_{Z^{\prime}}$ is the mass of $Z^{\prime}$-boson;
$\delta_{ZZ^{\prime}}=M_{Z^{\prime}}^2-m_{Z}^2$. The correction
$\Delta$ reflects the contribution from loop diagrams involving
all the particles that couple to the Higgs bosons [14,15]
\begin{eqnarray}
\label{e11}
\Delta=\left (\frac{\sqrt
N_{c}\,g\,m_{t}^2}{4\,\pi\,m_{W}\,\sin\beta}\right
)^2\,\log\left (\frac{m_{{\tilde t}_{1}}\cdot m_{{\tilde
t}_{2}}}{m_{t}^2}\right )^2\ ,
\end{eqnarray}
where $N_{c}$ is the number of colors, $m_{t}$ and $m_{W}$ are
the masses of top quark and $W$-boson, respectively. $\tan\beta$
defines the structure of the MSSM. The values of $\Delta\sim {\cal O}(
0.01~TeV^2)$ have been calculated [14] for any choice of
parameter space of the MSSM. We suggest that the
measurement of $M_{Z^{\prime}}$ would predict
the masses of mass-eigenstates ${\tilde t}_{1}$ and  ${\tilde
t}_{2}$,
since $m_{t}$ and $m_{W}$ are already measured in the
experiments and
$m_{h}$ is restricted by the LEP 2 data [1] as $m_{h}<$
130 GeV [16]; $M_{A}$ and $M_{H}$ are free parameters bounded by
combined data coming from the MSSM parameters space and the
experimental data [17].

Comparing (\ref{e8}) and (\ref{e9}), one can get the following
relation between $M_{Z^{\prime}}$ and $\Lambda$ as [9]
\begin{eqnarray}
\label{e12}
M_{Z^{\prime}}=\sqrt{\alpha_{em}}\,\Lambda\,\cot\phi/(2\,\sin\theta)\
,
\end{eqnarray}
where the value of $\Lambda $  was constrained from the CDF data at
 $\sqrt s$=1.8 TeV as
 $\Lambda >$ 3.7 TeV or 4.1 TeV, depending on the contact
interactions for the
left-handed electrons or muons, respectively found at 95 $\%$
confidence level [2,9].

In the decoupling regime of the MSSM Higgs sector where the couplings
of the light CP-even Higgs boson $h$ in the MSSM are identical
to those of the SM Higgs bosons and thus, the
CP-even mixing angle $\alpha$ behaves  as $\tan\alpha\rightarrow
-\cot\beta$ with the $M_{A}\gg m_{Z}$ relation, one can get $M_{H}^2\simeq
M_{A}^2+m_{Z}^2\sin^{2}(2\beta)+\mu^2$ which leads to disappearance
of the $H$-Higgs boson mass in (\ref{e10}). Here,
$\mu$ is the positive massive parameter which can, in principle, be
defined from the experiment searching for separation of two
degenerate heavy Higgs bosons, $A$ and $H$. This behavior
verified at the tree-level continues to hold even when radiative
corrections are included. It has been checked that this
decoupling regime is an effective one for all values of
$\tan\beta$ and that the pattern of most of the Higgs couplings
results from this limit.

In studying the mass relation (\ref{e10}) from the extended
electroweak gauge structure, we must be aware of the issues
related to the structure of $M_{Z^{\prime}}$ in both sides of
(\ref{e10}). We suppose that $M_{Z^{\prime}}$ in the l.h.s.
of (\ref{e10}) is the mass to be determined
using the CDF analysis data [2,9]. Therefore, one can
approximate its mass via the phenomenological relation (\ref{e12})
while the r.h.s. of (\ref{e10}) is model-dependent, where, to the
leading order, the mass $M_{Z^{\prime}}$ in the extended weak
interaction model  is  $ M_{Z^{\prime}}=m_{W}\,\sqrt{x}/(\cos\phi\sin\phi)$
[9] in the region where $\cos\phi< \sin\phi$.
With the help of the CDF restriction for $\Lambda$ [2] entering
into (\ref{e12}), one can easily find the upper limit
of the product of $m_{{\tilde t}_{1}}\cdot m_{{\tilde t}_{2}}$
from the following relation [13]
\begin{eqnarray}
\label{e13}
\Delta
<(B+M_{H}^{\star})\,(B-f\,C)+m_{h}^2-m_{Z}^2\,(1-\sin^{2}2\beta)+\mu^2\,
,
\end{eqnarray}
where $M_{H}^{\star}=(M_{A}^{2}+m_{Z}^2\,\sin^{2}2\beta
+\mu^2)^{1/2}$,
$f\equiv f(\phi)=\cot\phi\,\sqrt{\alpha_{em}}/(2\,\sin\theta)$,
$B\equiv B(x,\phi)=m_{W}\sqrt{x}/(\cos\phi\,\sin\phi)$, and $C$ is a
minimal value of the $\Lambda$ scale extracted from the CDF
analysis [2].
The masses of $Z$- and $W$-bosons are currently known with errors
of a few MeV each [18], whereas the mass of the top quark is known
with errors of a few GeV [18].
Note that the dependence of particle
couplings via $\tan\beta$ enters into the radiative correction
$\Delta$ in (\ref{e11}) and the mass $M_{H}$
defined in the decoupling regime. Thus,
the upper limit on $m_{{\tilde t}_{1}}\cdot m_{{\tilde t}_{2}}$
can be accurately predicted by precision measurements of the lower
bound of $M_{Z^{\prime}}$ and the masses of the Higgs bosons $h$ and $A$.

Fig. 1 shows the upper limit on
$L\equiv \log\left (\frac{m_{{\tilde t}_{1}}\cdot m_{{\tilde
t}_{2}}}{m_{t}^2}\right )$ as a function of $\sin\phi$ for $x=2$ and
$x=3$
at fixed values of  $\mu$ and $M_{A}$.
We use the following range of the mixing parameter
$0.75\leq\sin\phi\leq 0.85$ for a $Z^{\prime}$ boson [9] where
the luminosity required to exclude SU(2) $Z^{\prime}$ bosons of
various masses is lowest. The corresponding range of the lower
bound on $x$ is $3.9\geq x\geq 1.6$ (left-handed muons and
up-type quarks is taken into account) for the $\sin\phi$ range
above mentioned.
The regions of the parameter space lying below a
given line are allowed by the present model.
At present, the LEP bounds on the mass of
$A$-Higgs boson are $M_{A}>$ 88.4 GeV [1]. This result
corresponds to the large $\tan\beta$ region.
We see that the function $L$ is rather sensitive within the
changing of $\sin\phi$, i.e. the ratio of gauge couplings
$g/g_{l}$. Here, $g^{-2}=g_{l}^{-2}+g_{h}^{-2}$, where $g_{l}$
is associated with the $SU(2)_{l}$ group and defines the
couplings to the first and second generation fermions, whose
charges under subgroup $SU(2)_{l}$ are the same as in the SM, while
$g_{h}$ is originated from the $SU(2)_{h}$ group which governs
the weak interactions for the third generation (heavy) fermions.
 In the range of $\sin\phi$ presented in the Fig. 1, the width
 $\Gamma_{Z^{\prime}}$ of the $Z^{\prime}$-boson
falls to a minimum in the neighborhood of $\sin\phi =0.8$ [9], due
to the decreasing couplings to first two generations of fermions.
In the range $\sin\phi > 0.8$, $\Gamma_{Z^{\prime}}$
grows large, due to the rapid growth in the third generation
coupling.


\begin{center}
\begin{figure}[h]

\resizebox{12cm}{!}{\includegraphics{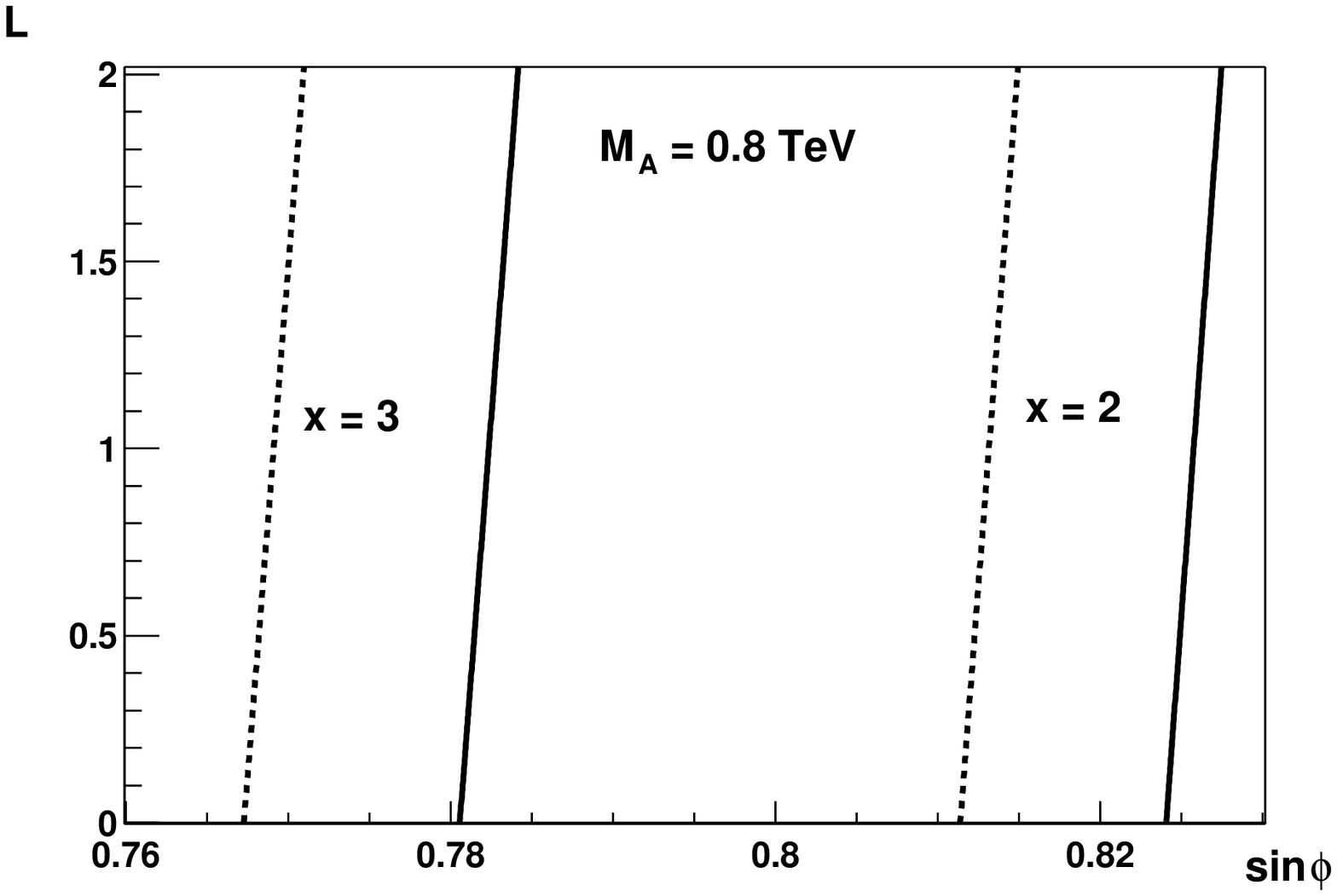}}

Fig. 1 The upper limit on
$L\equiv \log\left (\frac{m_{{\tilde t}_{1}}\cdot m_{{\tilde
t}_{2}}}{m_{t}^2}\right )$
as a function of $\sin\phi$ for different values of $x = 2$ and 3;
$\mu=m_{h}=120$~GeV (dashed line), $\mu=0$ (solid line) for
$M_A=0.8$~TeV; $\tan\beta$=30.

\end{figure}
\end{center}


\begin{center}
\begin{figure}[h]

\resizebox{12cm}{!}{\includegraphics{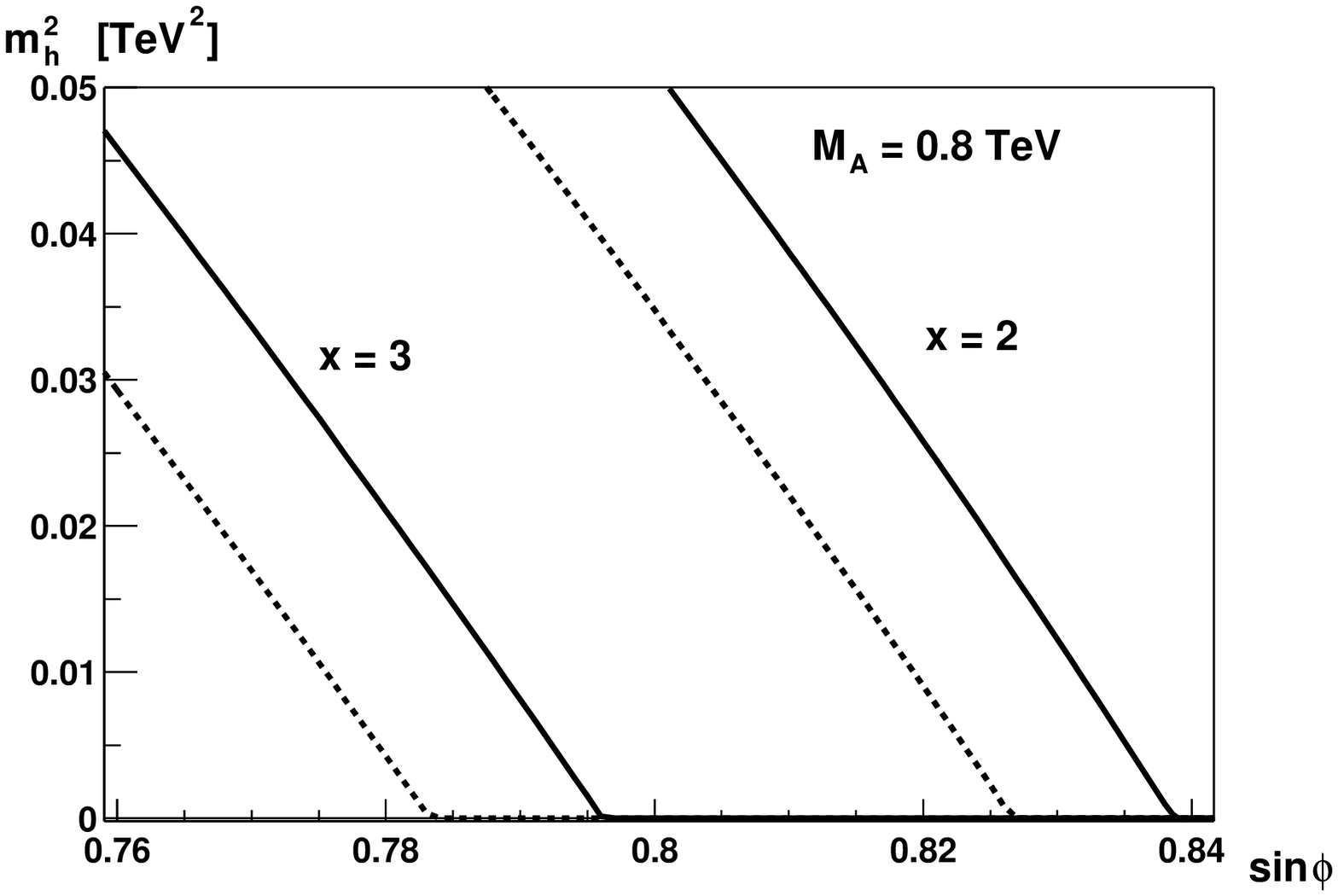}}

Fig. 2 The lower bound on $m^{2}_h$ as a function of $\sin\phi$
 for different values of $x = 2$ and 3;
$\mu=m_{h}=120$~GeV (dashed line), $\mu=0$ (solid line) for
$M_A=0.8$~TeV; $\tan\beta$=30.
 The regions of the parameter space lying
above a given line are allowed by the present model.

\end{figure}
\end{center}


The CDF analysis of the contact interaction between left-handed
muons and the up-type quarks is taken into account ($C=4.1$
TeV) in our calculations.
The lower bounds of $m^{2}_{h}$ are illustrated in Fig. 2.
The constraints are given for
different ratios of $x=2$ and $3$ as a function of
$\sin\phi$.
In our calculations, the parameters of the model are typically chosen as
 $m_{h}= 120$ GeV, $\tan\beta=30$.
The value of the neutral CP-odd Higgs boson $A$ mass is set to
be $M_{A}= 0.8$ TeV within the typical upper limit of $M_{A}$
kinematically allowed at the Tevatron Run II energy. Herewith,
we suppose that $A$ Higgs boson can be identified, e.g., via
two-lepton decays in the $AZ$ associated production process
$p\bar{p}\rightarrow AZ+X$ with $\sqrt{s} > M_{A}+m_{Z}$.
Here, we did not use the mass difference between $\tilde{t_{1}}$
and $\tilde{t_{2}}$ mass eigenstates, and we set
$m_{\tilde{t_{1}}}=m_{\tilde{t_{2}}}=$ 1 TeV (see Fig. 2).
The regions of the parameter space lying above a
given line are allowed by the present data.

In a more extended SUSY models, their mass sum rules can give
some useful estimations with the help of the CDF data [2].
For example, in the minimal $E_{6}$ superstring theory, the
particle spectrum consists of three scalar Higgs bosons $h$,
$H_{1}$, $H_{2}$, a pseudoscalar Higgs $A$, a charged Higgs
boson pair $H^{\pm}$, and two neutral gauge bosons $Z$ and
$Z^{\prime}$. Among these particles, there exists a mass
 sum rule, at the
tree-level, of the form  [19]:
\begin{eqnarray}
\label{e14}
M_{Z^{\prime}}^{2}=m_{h}^{2}+M_{H_{1}}^{2}+M_{H_{2}}^{2}-M_{A}^{2}-m_{Z}^{2}\,.
\end{eqnarray}
The analytical expressions for the loop corrections are unknown
yet. However, it is known that the one-loop corrections can be summarized into
the term logarithmically dependent on the SUSY sector mass scale
[19]. Considering that $m_{h}$ can be identified with
the lower bound on the Higgs boson mass, we obtain the
lower bound on the sum $M_{H_{1}}^{2}+M_{H_{2}}^{2}$ at fixed
$M_{A}$ as a function of $\sin\phi$:
\begin{eqnarray}
\label{e15}
\sum_{j=1}^{2} M_{H_{j}}^{2}> M_{A}^{2}+m_{Z}^{2}-m_{h}^{2}+
\frac{\alpha_{em}\,C^2\,\cot^{2}\phi}{4\,\sin^{2}\theta} \,.
\end{eqnarray}

The results of the calculation of the lower bound on $M\equiv (\sum_{j=1}^{2}
M_{H_{j}}^{2})^{1/2}$  as the function of $\sin\phi$ is given in the Fig. 3
\begin{center}
\begin{figure}[h]

\resizebox{12cm}{!}{\includegraphics{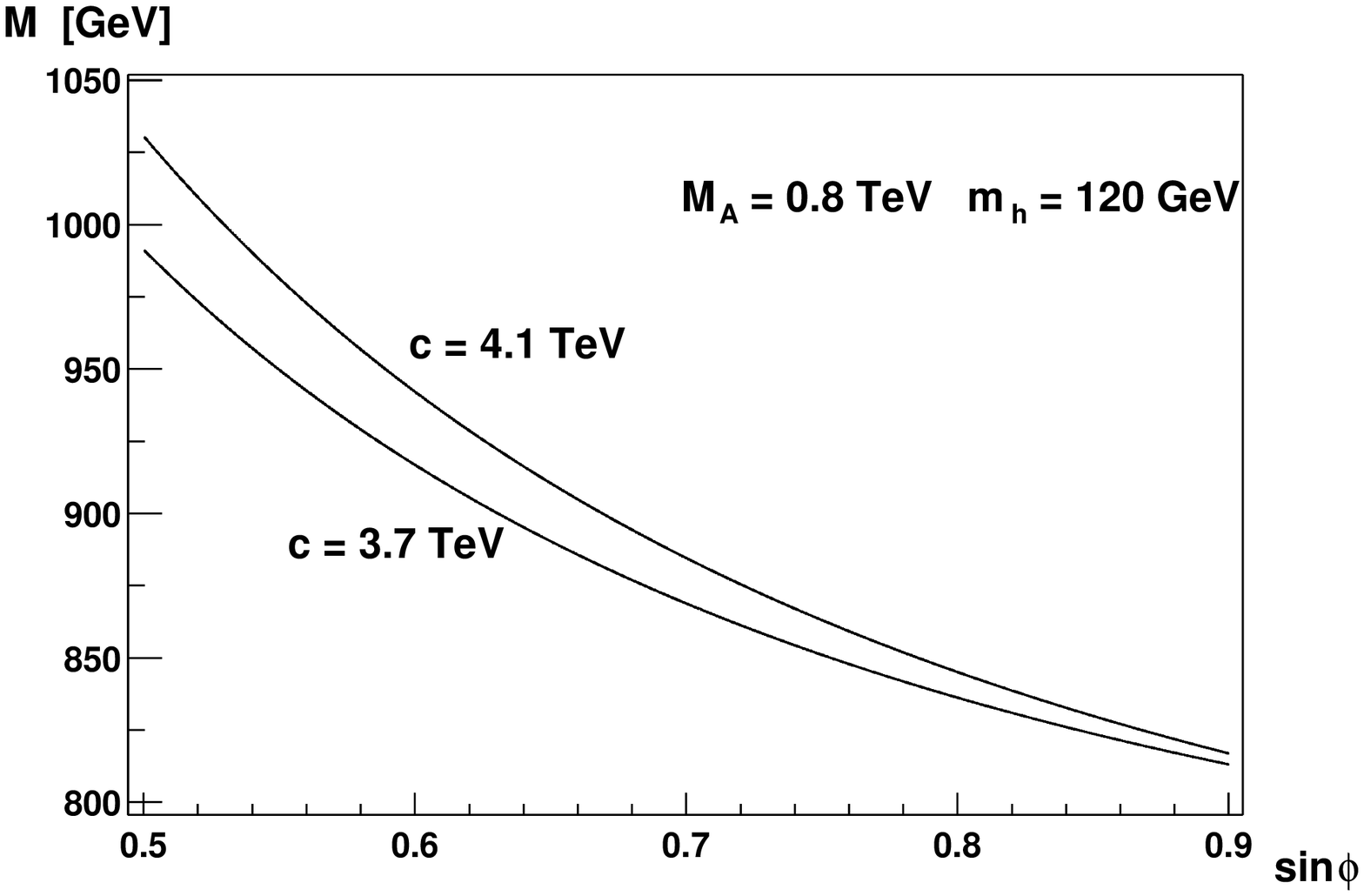}}

Fig. 3 The lower bound on $M\equiv (\sum_{j=1}^{2}
M_{H_{j}}^{2})^{1/2}$  as the function of $\sin\phi$.

\end{figure}
\end{center}

We have used the scales of new physics $\Lambda >C$ coming from the
CDF analysis [9] at 95 $\%$ confidence level
in which contact interactions were assumed only between left-handed
electron (muon) and up-type quarks: $\Lambda> 4.1$ TeV
and  $\Lambda > 3.7$ TeV for left-handed muons  and
 left-handed electrons, respectively.

\section{Rare decays of the Higgs bosons}

As is well known, Higgs bosons dominantly couple to heavy particles
even in the MSSM.
 Observation
of $h$-, $H$- and $A$-Higgs bosons depends on the
model parameters including the
masses $m_{h}$, $m_{H}$ and $M_{A}$.

(a)  The decays $h,H\rightarrow gg, \gamma\gamma$.

Let us begin with
 study on the decays
$X\rightarrow gg$ and $X\rightarrow \gamma\gamma$ ($X:h,H$), where the
former dominates over the later.
The decay width of $X\rightarrow gg$ (the radiative
corrections are not included) is given by
\begin{eqnarray}
\label{e16}
\Gamma (X\rightarrow gg)=
\frac{g^2\,\alpha_{s}^{2}\,m_{X}^3}{128\,\pi^3\,m_{W}^2}\cdot {\vert\tilde
F_{Q}\vert }^2\ ,
\end{eqnarray}
where the  transition formfactor $\tilde F_{Q}$ is
provided by an intermediate quark loop containing $b$- and
$t$-quarks, [20,21]:
\begin{eqnarray}
\label{e17}
\tilde
F_{Q}=\rho_{Xb}\,\tau_{b}\left\{\frac{(\tau_{b}-1)}{4}\left
[\pi^2-\ln^{2}\left (\frac{1+\xi_{b}}{1-\xi_{b}}\right )\right
]-1\right\}\cr
 +\rho_{Xt}\,\tau_{t}\left\{(\tau_{t}-1)\left [\arcsin^{2}\left
(\frac{m_{X}}{2\,m_{t}}\right )\cdot\theta (2\,m_{t}-m_{X}) \right.\right.\cr
\left.\left. +\frac{1}{4}\,\left [\pi^2-\ln^{2}
\left (\frac{1+\xi_{t}}{1-\xi_{t}}\right )\right
]\cdot\theta (m_{X}-2\,m_{t})\right ] -1\right\}\ .
\end{eqnarray}
Here, $\tau_{Q}=(2\,m_{Q}/m_{X})^2$, $\xi_{Q}=\sqrt
{1-\tau_{Q}}$, $\rho_{hQ}=-\sin\alpha/\cos\beta$ and
$\rho_{HQ}=\cos\alpha/\cos\beta$ for b-quarks;
$\rho_{hQ}=\cos\alpha/\sin\beta$ and
$\rho_{HQ}=\sin\alpha/\sin\beta$ for t-quarks; the mixing angle $\alpha$
diagonalizes the CP-even Higgs squared-mass matrix. For comparison,
the decay width $\Gamma (X\rightarrow\gamma\gamma)$ can be
easily obtained from the corresponding decay to two gluons
(\ref{e16}) by a simple replacement
$\alpha_{s}\rightarrow\frac{3}{\sqrt 2}\alpha\,e_{Q}^2$
(the heavy quark charges $e_{Q}$ are included into the formfactors)
and $\tilde F_{Q}\rightarrow F_{Q}+F_{W}+F^{\pm}_{H}$
\begin{eqnarray}
\label{e18}
\Gamma (X\rightarrow\gamma\gamma)=
\frac{g^2\,\alpha^2 \,m_{X}^3}{1024\,\pi^3\,m_{W}^2}\cdot {\vert
F_{Q}+F_{W}+ F_{H^{\pm}}\vert }^2\ ,
\end{eqnarray}
where the formfactor $F_{Q}$ contributed from heavy quarks $Q$ is given by
\begin{eqnarray}
\label{e19}
F_{Q}=2\left \{\rho_{Xb}\frac{\tau_{b}}{3}\left [\frac{(\tau_{b}-1)}{4}\left
(\pi^2-\ln^{2}\frac{1+\xi_{b}}{1-\xi_{b}}\right )-1\right ]\right.\cr
\left. +\frac{4}{3}\,\rho_{Xt}\,\tau_{t}\left [(\tau_{t}-1)\left (\arcsin^{2}\left
(\frac{m_{X}}{2\,m_{t}}\right )\cdot\theta (2\,m_{t}-m_{X})\right. \right.\right.\cr
\left.\left.\left. +\frac{1}{4}\,\left (\pi^2-\ln^{2}\frac{1+\xi_{t}}{1-\xi_{t}}\right
)\cdot\theta (m_{X}-2\,m_{t})\right )\right ] -1\right\}\
\end{eqnarray}
and the ones $F_{W}$ from weak bosons $W^{\pm}$  and $ F_{H^{\pm}}$ from charged Higgs
bosons $H^{\pm}$ are [22]
\begin{eqnarray}
\label{e20}
F_{W}=\rho_{XW}\left \{2+3\,\tau_{W}+3\,\tau_{W}(2-\tau_{W})\left
[\arcsin^{2}\left (\frac{m_{X}}{2\,m_{W}}\right )\cdot\theta (2\,m_{W}-m_{X})
 \right.\right.\cr
\left.\left. +\frac{1}{4}\,\left (\pi^2-\ln^{2}
\frac{1+\xi_{W}}{1-\xi_{W}}\right )
\cdot\theta (m_{X}-2\,m_{W})\right ]\right \}
\end{eqnarray}
and
\begin{eqnarray}
\label{e21}
F_{H^{\pm}}=\rho_{XH^{\pm}}\left (\frac{m_{W}}{M_{H^{\pm}}}\right
)^2\left
(1-\tau_{H^{\pm}}\,\arcsin^{2}\frac{m_{X}}{2\,M_{H^{\pm}}}\right
)\ ,
\end{eqnarray}
respectively. Here, $\rho_{hW}=\sin(\beta-\alpha)$,
$\rho_{HW}=\cos(\beta-\alpha)$,
 $\rho_{hH^{\pm}}=\rho_{hW}+
\frac{\cos 2\beta\,\sin(\beta+\alpha)}{2\,\cos^{2}\theta_{W}}$,
$\tau_{H^{\pm}}=(2\,M_{H^{\pm}}/m_{X})^2$.
Calculated results of
$\Gamma (H\rightarrow gg)$,  $\Gamma (h\rightarrow gg)$ and
$\Gamma (H\rightarrow\gamma\gamma)$, $\Gamma (h\rightarrow\gamma\gamma)$
are given in the Figs. 4, 5 and 6, 7, respectively, where
 the mass of the CP-odd $A$-Higgs boson was typically taken to be $M_{A}=$ 0.8 TeV.
\begin{center}
\begin{figure}[h]
\resizebox{14cm}{!}{\includegraphics{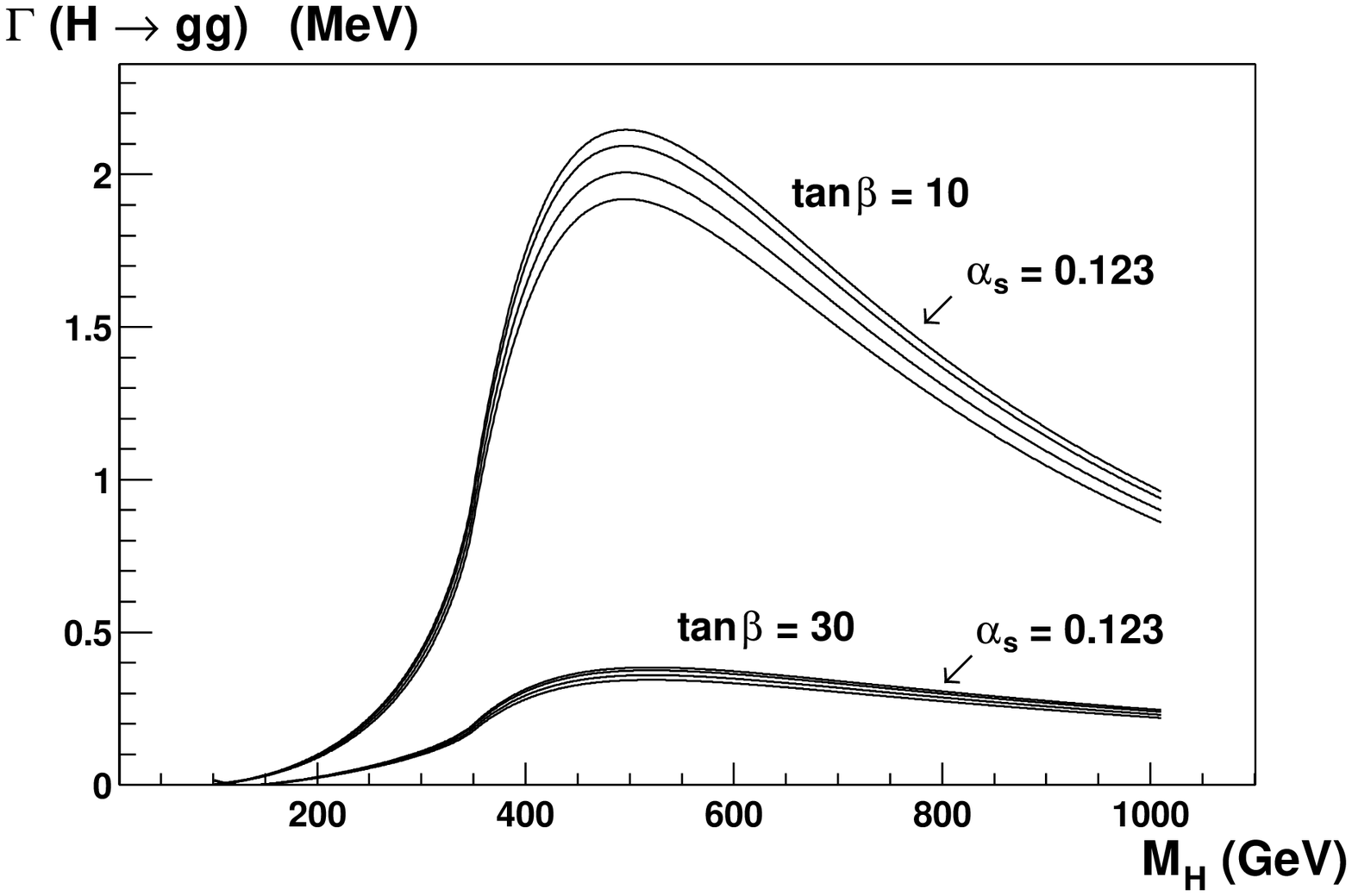}}

Fig. 4  The  decay width $\Gamma (H\rightarrow gg)$ as the
function of $M_{H}$ for different $\alpha_{s}$ (from below $\alpha_{s}$
= 0.110, 0.115, 0.119, 0.123) and $\tan\beta$= 10 and 30.
\end{figure}
\end{center}
\begin{center}
\begin{figure}[h]
\resizebox{14cm}{!}{\includegraphics{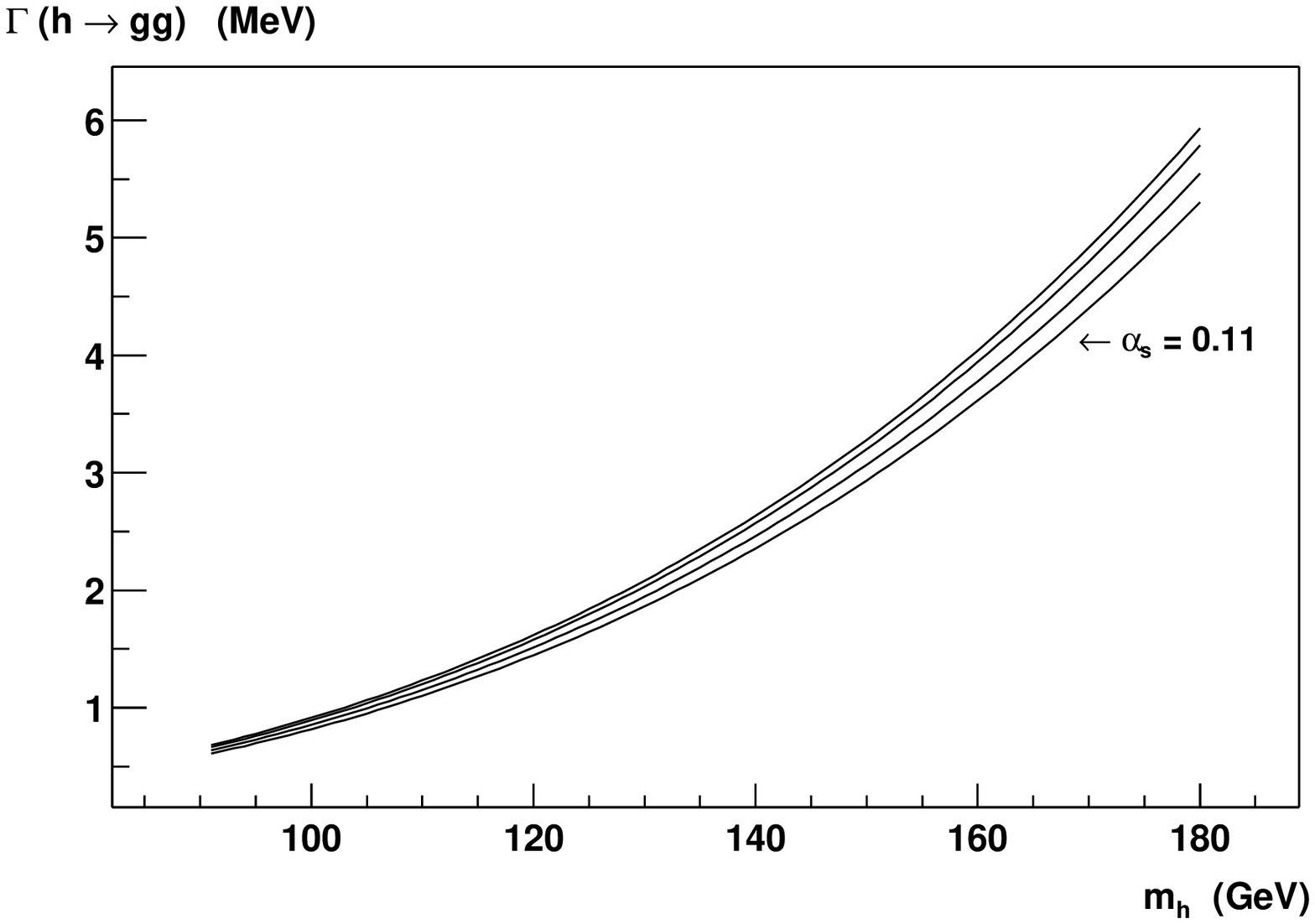}}

Fig. 5  The  decay width $\Gamma (h\rightarrow gg)$ as the
function of $m_{h}$ for different $\alpha_{s}$ (from below $\alpha_{s}$
= 0.110, 0.115, 0.119, 0.123) in the decoupling limit.
\end{figure}
\end{center}
\begin{center}
\begin{figure}[h]
\resizebox{14cm}{!}{\includegraphics{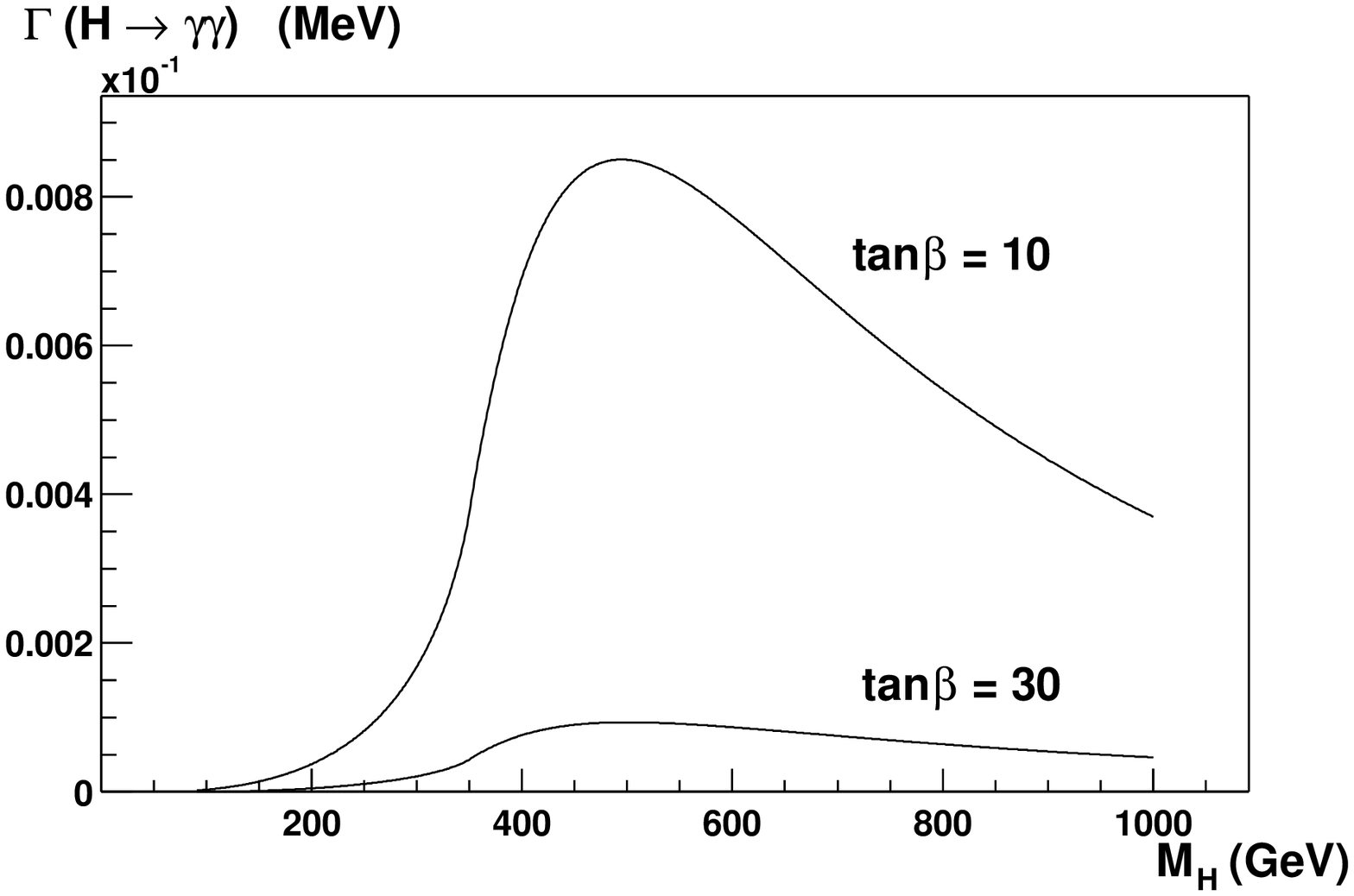}}

Fig. 6  The  decay width $\Gamma (H\rightarrow\gamma\gamma)$ as the
function of $M_{H}$ for different $\tan\beta$= 10 and 30.
\end{figure}
\end{center}
\begin{center}
\begin{figure}[h]
\resizebox{14cm}{!}{\includegraphics{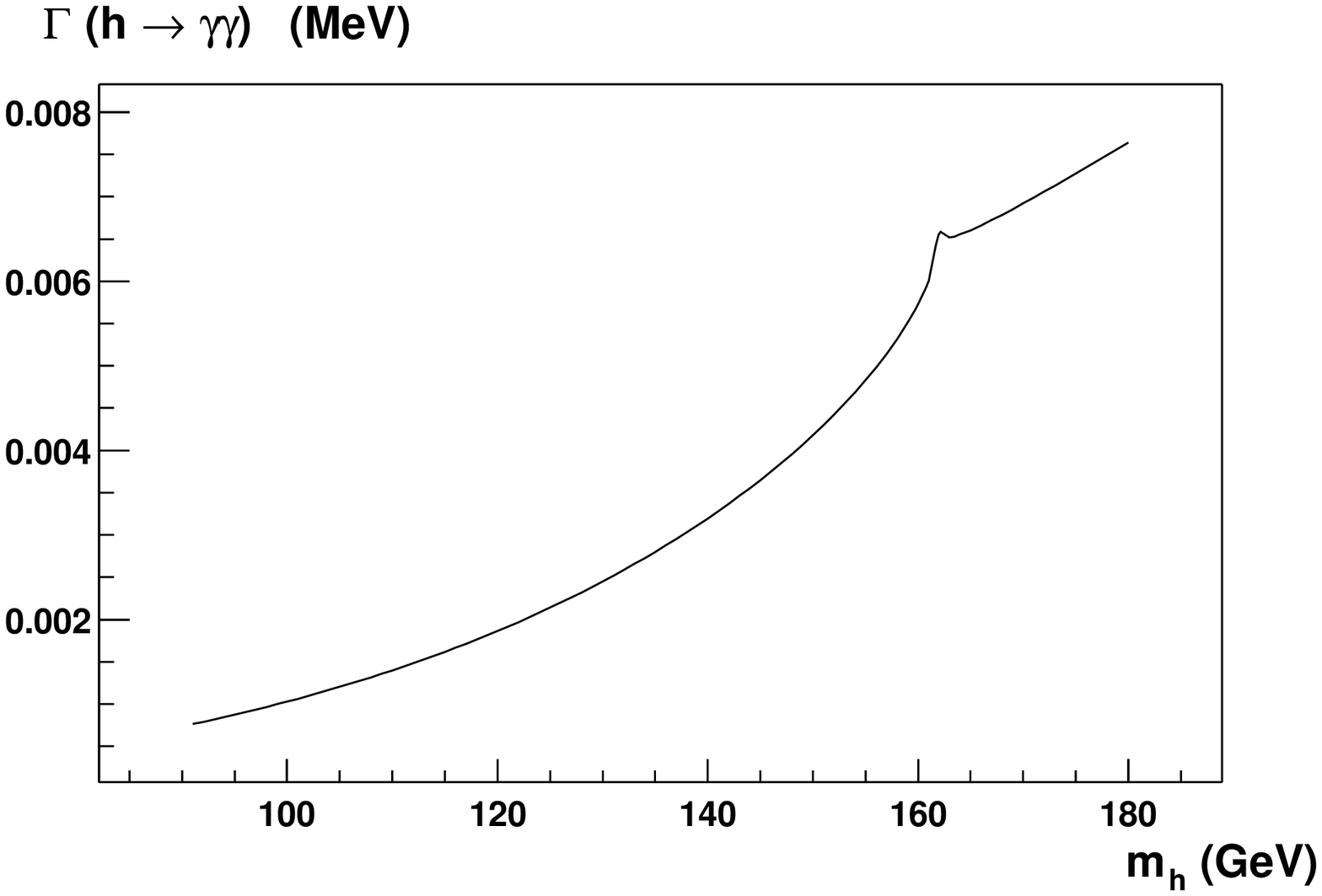}}

Fig. 7  The  decay width $\Gamma (h\rightarrow\gamma\gamma)$ as the
function of $m_{h}$ in the decoupling limit.
\end{figure}
\end{center}
In our calculations, the couplings between $h$-Higgs and left-and
 right-handed
sfermions as well as $h$-Higgs to charginos were neglected.
 The following relation between the
mixing angle $\alpha$ and $\tan\beta$
\begin{eqnarray}
\label{e22}
\cot\alpha=-\tan\beta\left [1+2{\left
(\frac{m_{Z}}{M_{A}}\right )}^2\,\cos 2\beta\right ] + {\cal O}\left
(\frac{m^{4}_{Z}}{M^{4}_{A}}\right )
\end{eqnarray}
is useful in evaluating $h$-Higgs boson couplings to $b$-and $t$-quarks
in the decoupling regime.

 Here, we point out
that the Run II at the Tevatron could observe the Higgs boson $h$
through the promising final states of $gg$ and
$\gamma\gamma$ channels or $\bar{b}b$ final state.\\

(b) The decays $A\rightarrow \bar{l}l$.

In the MSSM, the decay amplitude of a CP-odd neutral Higgs boson $A$ (with
mass $M_{A}$ and 4-momentum $P_{\mu}$) into a  lepton $l$
and antilepton $\bar{l}$ with 4-momenta $p_{\mu}$ and
$(P-p)_{\mu}$, respectively, is given by the following Feynman
amplitude
\begin{eqnarray}
\label{e23}
Am(A\rightarrow\bar{l}l)=i\,\sum_{Q:b,t}\bar{u}(p)\,\gamma_{5}\,F_{Q}(M_{A}^2)\,
v(P-p)\ ,
\end{eqnarray}
where $F_{Q}(t=M_{A}^2)$ is the complex function (formfactor) providing the
transition from the $A$ Higgs boson into a lepton pair. One can
consider this transition via the intermediate of-shell spin-1 bosons $V^{*}$ (photons,
$W^{\pm}$-bosons, and so on) within the vertex $AV^{*}V^{*}$
\begin{eqnarray}
\label{e24}
\Gamma_{\mu\nu}=\epsilon_{\mu\nu\alpha\beta}\,k^{\alpha}\,(P-k)^{\beta}\,F_{A}[
k^2,(P-k)^{2}]\
\end{eqnarray}
with
\begin{eqnarray}
\label{e25}
F_{A}[k^2,(P-k)^{2}]=\sum_{Q:b,t}\rho_{Q}\,e_{Q}^2\,f_{VV}^{Q}
\,\hat{F_{QA}}[k^2,(P-k)^{2}]\
.
\end{eqnarray}
Here, $\rho_{Q}=\tan\beta$ for $b$-quark-loop and $\rho_{Q}=\cot\beta$ for
$t$-quark-loop; $k_{\mu}$ is the
4-momentum of a $V^{*}$ -boson. There is a natural normalization
condition $F_{A}(0,0)$=$f_{VV}$ in (\ref{e24}) where the decay constant
$f_{VV}$ is given for the decay of $A$-Higgs into two real $V$
-bosons. The form of $F_{A}$ in (\ref{e25}) should be taken so
that the Feynman integrals are not divergent. Here, we simply take
 $\hat{F_{QA}}[k^2,(P-k)^{2}]$ as
\begin{eqnarray}
\label{e26}
\hat{F_{QA}}[k^2,(P-k)^{2}]=\frac{\lambda_{Q}^2}{\lambda_{Q}^2-k^2-(P-k)^{2}}
\end{eqnarray}
which has the good analytical properties, and carries both the
vector dominance features and the properties of the right static
limit. The parameter $\lambda_{Q}$ must be large enough, and
for numerical estimations we put $\lambda_{Q}^2=M_{A}^2$ where
$M_{A}\sim {\cal O}$(1 TeV). To be more precise in the calculations of
the decay width of the process $A\rightarrow\bar{l}l$, one has
to take into account the fact that the formfactor $F_{A}[k^2,(P-k)^{2}]$
should be complex function with the following real part
\begin{eqnarray}
\label{e27}
Re~ F_{A}(M_{A}^2)=\frac{P}{\pi}\int_{0}^{\infty}\frac{Abs
[F_{A}(t)]}{t-M_{A}^2}\,dt\ .
\end{eqnarray}
Here, the absorptive part $Abs [F_{A}(t)]$ can be found if we take
the $S_{0}^{1}$ leptonic pair in the final state and taking into account
the unitarity condition for the decay amplitude
\begin{eqnarray}
\label{e28}
Abs [F_{A}(t)]=\frac{1}{2\,i\,\sqrt{2\,t}}\int
d\Omega_{I}\,\delta^{4}(P-\sum_{I} q_{I}){\langle I\vert T\vert
(\bar{l}l)_{S_{0}^{1}}\rangle}^{*}\,Am (A\rightarrow I)\ .
\end{eqnarray}
In (\ref{e28}), the integration over the phase-space
volume $\Omega_{I}$ for all possible intermediate states $I$, i.e.
off-shell $\gamma^{*}\gamma^{*}$-state, heavy quark-antiquark
state, heavy quark and the of-shell $\gamma^{*}$-quantum are taken
into account.
The decay width $\Gamma(A\rightarrow\bar{l}l)$ normalized to
two photon decays is calculated as follows [20]
\begin{eqnarray}
\label{e29}
Br(A\rightarrow\bar{l}l/\gamma\gamma)\equiv\frac{\Gamma(A\rightarrow\bar{l}l)}
{\Gamma(A\rightarrow\gamma\gamma)}=2\,\xi_{l}\,\frac{1}{(4\,\pi)^2}\,\left
(\frac{m_{l}}{M_{A}}\right )^2\frac{1}{\pi^2}\,{\vert R\vert}^2\
\end{eqnarray}
with
\begin{eqnarray}
\label{e30}
R=\frac{1}{\xi_{l}}\left [\frac{i\,\pi}{2}\ln\frac{1-\xi_{l}}{1+\xi_{l}}+
\frac{1}{4}\ln^{2}\frac{1+\xi_{l}}{1-\xi_{l}}-\ln\frac{1+\xi_{l}}{1-\xi_{l}}
+\frac{\pi^2}{12}-\Phi\left (\frac{1-\xi_{l}}{1+\xi_{l}}\right
)\right ]
  ,
\end{eqnarray}
where $\Phi(x)=\int_{0}^{x}\,dt\,\ln(1+t)/t$,
$\xi_{l}=\sqrt{1-(2\,m_{l}/M_{A})^2}$.
The normalized decay rate (\ref{e29}) is very convenient because the
dependence on $\tan\beta$ is cancelled for the $VV$-channel.
The corresponding SM background comes from the Drell-Yan
production
$$\bar{q}q\rightarrow\gamma^{*},Z^{*},Z^{\prime}{^*}\rightarrow\bar{l}l\
.$$
The signals we are looking for would be identified by the
peaks of the invariant mass of
 $\tau^{+}\tau^{-}$ and/or $\mu^{+}\mu^{-}$ pairs.
 In $Table~ 1$ we show the calculated results of the decay widths
$A\rightarrow\mu^{+}\mu^{-}$ and $A\rightarrow\tau^{+}\tau^{-}$
normalized to $\gamma\gamma$- channel as the function of $M_{A}$.

{\bf $\underline {Table 1}$. } The values
$Br(A\rightarrow\bar{l}l/\gamma\gamma)$ as the function of
$M_{A}$ where $l=\mu^{-}$ and $\tau^{-}$.

\begin{center}
\begin{tabular}{c c c c c c c  } \hline
\multicolumn{7}{c}
{$M_{A} (TeV)$} \\[0.5 mm]
\hline
$ $& 0.10 & 0.25 & 0.50 & 0.75 & 1.00 & 1.20  \\[0.5 mm]
\hline
$ Br(A\rightarrow\mu^{+}\mu^{-}/\gamma\gamma)\cdot 10^{8}$ & 3.00 & 1.00 &0.28 & 0.15 & 0.10 & 0.007  \\[0.5 mm]
$ Br(A\rightarrow\tau^{+}\tau^{-}/\gamma\gamma)\cdot 10^{6}$ & 0.84 & 0.27 & 0.11 & 0.06 & 0.04 & 0.02  \\[0.5 mm]
\hline
\end{tabular}
\end{center}
At the end of this part, we conclude that apart from reducing
of the
$A\rightarrow\tau^{+}\tau^{-}$-signal due to some
experimental constraints, this decay mode could be
detected at the Tevatron's Run II as (almost) easily as the corresponding signal of
$A\rightarrow\mu^{+}\mu^{-}$ decay.\\

(c) The decays $A\rightarrow\gamma\gamma h, ggh$.

In addition to the proposal for searching for $A$-Higgs bosons via
the decay $A\rightarrow Zh$ [10], we suggest to investigate the
decays $A\rightarrow\gamma\gamma h$ and $A\rightarrow ggh$ which can be
relevant at the Tevatron's Run II for $M_{A}>m_{h}$ at large $\tan\beta$.
We first note that the rate of the $\gamma\gamma h$ channel can be more
promising for discovery  for $A$-Higgs bosons than that of
$\tau^{+}\tau^{-}$ or $\mu^{+}\mu^{-}$ channels.

The matrix element of the decay of $A$-Higgs bosons with momentum $P$
into $\gamma$, $\gamma$ and $h$ with the momenta $k_{1}, k_{2}$ and
$k_{3}$, respectively, has the following form
\begin{eqnarray}
\label{e31}
M(A\rightarrow\gamma\gamma
h)=\frac{\sqrt{N_{c}}\,\pi\,\alpha\,i\,g}{m_{W}\,\sqrt{M_{A}}}\,
Tr \left [\sum_{Q:b,t}
\gamma_{5}\,\rho_{Q}\,e_{Q}^{2}\,m_{Q}\,(M_{A}-\hat{P})
\Gamma_{Q}\,\rho_{hQ}\,T_{Q}\right ]\ ,
\end{eqnarray}
where $T_{Q}$ is the amplitude providing the transition of the
$\bar{Q}Q$-virtual quark pair into the $\gamma\gamma h$ final state
(the permutations are taken into account); $\Gamma_{Q}$ is the
vertex function describing the couplings $h\bar{Q}Q$
taking into account the
$\rho_{Q}$ flavor-dependent factor.
The differential distribution of the decay width
 $A\rightarrow\gamma\gamma h$ over the invariant mass $\tilde s$ of
final state's two-photon pairs and normalized to the
$\gamma\gamma$ width is given by
\begin{eqnarray}
\label{e32}
\frac{1}{\Gamma(A\rightarrow\gamma\gamma)}
\frac{d \Gamma(A\rightarrow\gamma\gamma h)}{d \tilde{s}}=
\rho^{2}_{hQ}\frac{G_{F}\,a^{3}\,\tilde{s}\,M_{A}^{2}}
{8\,\sqrt{2}\,\pi^{2}\,(a^2-{\tilde{s}}^2)^2\,(a+\tilde{s}/2)},
\end{eqnarray}
where $a=(1-\kappa^2)/2$, $\kappa=m_{h}/M_{A}$,
$\tilde{s}=s/M_{A}^2$, $s=(k_{1}+k_{2})^2$. In $Table~ 2$, we
present $Br(A\rightarrow\gamma\gamma h/\gamma\gamma)$ as a
function of $\kappa$ for different values of $\tan\beta=5,10,20$
and $50$ at fixed $M_{A}$=0.8 TeV.

{\bf $\underline {Table~2}$. } $Br(A\rightarrow\gamma\gamma h/\gamma\gamma)\cdot
10^{-2}$ as a
function of $\kappa$ for $\tan\beta=5,10,20$ and $50$;
$M_{A}$ is taken to be $M_{A}$=0.8 TeV as a typical value.

\begin{center}
\begin{tabular}{ c c c c c c c  } \hline
\multicolumn{7}{c}
{$\kappa=m_{h}/M_{A}$} \\[0.5 mm]
\hline
$ $& 0.05 & 0.1 & 0.2 & 0.3 & 0.4 & 0.5  \\[0.5 mm]
\hline
$ \tan\beta =5$ & 0.066  & 0.054  & 0.044 & 0.032  & 0.025  & 0.014  \\[0.5 mm]
$ \tan\beta =10$ & 0.26 & 0.22 & 0.18 & 0.13 & 0.10 & 0.058  \\[0.5 mm]
$ \tan\beta =20$ & 1.09 & 0.91 & 0.73 & 0.54 & 0.42 & 0.24  \\[0.5 mm]
$\tan\beta =50 $ &  6.54 & 5.43 & 4.43 & 3.28& 2.54 & 1.43 \\[0.5 mm]
\hline
\end{tabular}
\end{center}
The width $\Gamma (A\rightarrow ggh)$ can be obtained from the
corresponding decay width for $A\rightarrow \gamma\gamma h$ by
making a replacement
$\alpha\,e_{Q}^2\rightarrow (\sqrt{2}/3)\,\alpha_{s}$. The
relative decay width of $A\rightarrow ggh$ normalized to
$A\rightarrow gg$ gives an identical value to the one listed in
 $Table~ 2$ at  $M_{A}$=0.8 TeV.

Compared with
$A\rightarrow\tau^{+}\tau^{-}$/$\mu^{+}\mu^{-}$ cases as discussed
in (b), we conclude that the $A\rightarrow\gamma\gamma
h$ channel in MSSM could be observed for 0.15$\leq\kappa\leq$0.2
(120 GeV$\leq m_{h}\leq$ 160 GeV) and for $\tan\beta >$5.

\section{The decays $Z_{2}\rightarrow Z_{1}h$}
To reach a unified theory of all interactions one could start with
Grand Unification Theory (GUT) group
$SU(5)$ [23] which is
the minimal unification group of strong and electromagnetic
interactions. However, this example was ruled out by several
experiments such as searching for the proton decay. The
natural extension leads to $SO(10)$ [24]. It is known that all
unification groups larger than $SU(5)$ have extra gauge bosons,
 the neutral $Z^{\prime}$ bosons and the charged ${W^{\pm}}^{\prime}$ ones.
Experimental signals of those extra gauge bosons would give very
important information about the underlying GUT and its origin.
Search for $Z^{\prime}$ and ${W^{\pm}}^{\prime}$ is therefore an
important subject for the physical program at the Tevatron's
Run II, where very high precision measurements
can give valuable information on $Z^{\prime}$ and ${W^{\pm}}^{\prime}$
because they are sensitive to rare processes including their
decays.

The cross-section $\sigma (\bar{p}p\rightarrow
Z^{\prime}\bar{l}l)$ for $Z^{\prime}$ production at the Tevatron
is inversely proportional to the total decay width
$\Gamma_{Z^{\prime}}$. Obviously, the inclusion of extra
channels to $Z^{\prime}$-decays leads to that $\Gamma_{Z^{\prime}}$
will become larger and $\sigma (\bar{p}p\rightarrow
Z^{\prime}\bar{l}l)$ smaller.
As already pointed out [25,26], if the decays of $Z^{\prime}$ into
$h +\{\bar{Q}Q\}_{s=1}$, $h\bar{l}l$ are
kinematically allowed, search for $h$ Higgs boson or an exotic heavy
quark-antiquark bound states $\{\bar{Q}Q\}_{s=1}$ with  spin
$s=1$ would be the interesting channels.
The study of the
decay $Z^{\prime}\rightarrow h +\{\bar{Q}Q\}_{s=1}$ [25,26]
would provide a useful information about the nature of the
extended gauge structure such as the couplings of $Z^{\prime}$
with heavy quarks in both vector and axial-vector sectors,
couplings of $h$ with heavy quarks, $Z-Z^{\prime}$ mixing effects,
physics of mass eigenstates $Z_{1}$ and $Z_{2}$ and
interplay with matter fields.

In an effective rank-5 model, including only one extra neutral
gauge boson $Z^{\prime}$, the interaction Lagrangian is given by
standard manner
\begin{eqnarray}
\label{e33}
-{\cal L}_{int}=\frac{1}{2}g_{1}Z_{\mu}\left
[\sum_{f}\bar\Psi_{f}\gamma^{\mu}
(g_{v}^{f}-g_{a}^{f}\gamma_{5} )\Psi_{f}\right ]\cr
+\frac{1}{2}g_{2}Z^{\prime}_{\mu}\left
[\sum_{f}\bar\Psi_{f}\gamma^{\mu}
(g_{v}^{f\,\prime}-g_{a}^{f\,\prime}\gamma_{5} )\Psi_{f}\right
]\, ,
\end{eqnarray}
where $\Psi_{f}$ is the fermion field with the flavor $f$. The
first term in (\ref{e33}) is written down within the SM where
$g_{v}^{f}=T_{3\,L}-2\,Q_{f}\,\sin^{2}\theta_{W}$,
$g_{a}^{f}=T_{3\,L}$; $g_{1}=g/\cos\theta_{W}$ is the SM
coupling constant, $T_{3\,L}$ and $Q_{f}$ are the third component
of the weak isospin and the electric charge, respectively. The
pairs $(g_{v}^{f},g_{a}^{f})$ and  $(g_{v}^{f\,\prime},g_{a}^{f\,\prime})$
represent the chiral properties of interactions of $Z$- and $Z^{\prime}$-bosons
to $\Psi_{f}$, respectively. The mass eigenstates
$Z_{1}$ and $Z_{2}$ are parameterized by a mixing angle
$\theta$ originated from the mixing between weak eigenstates
 $Z$- and $Z^{\prime}$
\begin{eqnarray}
\label{e34}
{Z_{1}\choose Z_{2}}=
\left(\matrix{\cos\theta&\sin\theta\cr
-\sin\theta&\cos\theta\cr}\right)\cdot {Z\choose Z^{\prime}}\ .
\end{eqnarray}
Therefore, the Lagrangian (\ref{e33}) is replaced by
\begin{eqnarray}
\label{e35}
{\cal L}_{int}=\frac{-g}{2\cos\theta_{W}}\sum_{f}\left
[Z_{1_{\mu}}\bar\Psi_{f}\gamma^{\mu}
(V^{f}-A^{f}\gamma_{5} )\Psi_{f}
+Z_{2_{\mu}}\bar\Psi_{f}\gamma^{\mu}
(V^{f\,\prime}-A^{f\,\prime}\gamma_{5} )\Psi_{f}\right],
\end{eqnarray}
where
\begin{eqnarray}
\label{e36}
V^{f}=g_{v}^{f}\cos\theta+\frac{g_{2}}{g_{1}}\,g_{v}^{f\,\prime}\,\sin\theta\
,\,\,A^{f}=g_{a}^{f}\cos\theta+\frac{g_{2}}{g_{1}}\,g_{a}^{f\,\prime}\,\sin\theta\,
\end{eqnarray}
and
\begin{eqnarray}
\label{e37}
V^{f\,\prime}=\frac{g_{2}}{g_{1}}\,g_{v}^{f\,\prime}\,\cos\theta-g_{v}^{f}\,\sin\theta\
,\,\,A^{f\,\prime}=\frac{g_{2}}{g_{1}}\,g_{a}^{f\,\prime}\,\cos\theta
-g_{a}^{f}\,\sin\theta\,
\end{eqnarray}
with $g_{2}=g_{1}\sqrt{(5/3)\,\sin^{2}\theta_{W}\,\lambda}\simeq
g_{1}\cdot 0.62\sqrt{\lambda}$, $\lambda\sim {\cal O}$(1) [27].
We use the LEP measured value $\sin^{2}\theta_{W}(\overline{MS})$=0.23117 [18].

The decays of $Z_{2}$ are the promising place to search for
CP-even light Higgs-boson $h$. Here, the effects of heavy quarks
cannot be neglected. As a result, there exist an effective
$h$-gluon-gluon interaction which arises from the triangle
diagram with a heavy quark loop and does not decouple in the
limit of large quark masses. One can separate out the heavy
quark contribution and use an effective low-energy theorem [28-30]
for Higgs boson interactions. In the limit $M_{Z_{2}}\gg m_{h}$
($m_{h}\sim {\cal O}$(100 GeV), $M_{Z_{2}}$ is the mass of $Z_{2}$)
when $h$ is a constant field, the interactions of $h$ is
reproduced by rescaling all the mass terms $m_{j}=m_{j}(1+h/v)$,
$j$= quarks, W, $Z, Z^{\prime}$,..., and $\alpha_{s}\rightarrow
\alpha_{s}+\delta\alpha_{s}$ with
 $\delta\alpha_{s}=\alpha_{s}^2\,h/(3\,\pi\,v)$ [28-30].
Here, the number of heavy quarks is restricted by one, that is only
top quark loop will be involved into the game.
The interaction  Lagrangian looks like:
\begin{eqnarray}
\label{e38}
{\cal L}_{int}=(1+h/v)\rho_{ht}\,m_{t}\,\bar{t}t+
\frac{\alpha_{s}\,\rho_{ht}}{12\,\pi\,v}\,h\,G_{\mu\nu}\,G^{\mu\nu}\cr
-(1+h/v)^2 \left (m_{W}^2
W_{\mu}^{+}W^{\mu\-}+\frac{1}{2} m_{Z}^2 Z_{\mu}Z^{\mu} +
\frac{1}{2} M_{Z^{\prime}}^2 Z_{\mu}^{\prime} Z^{\mu\,\prime}\right ) \,,
\end{eqnarray}
where $G_{\mu\nu}$ is the standard gluon field strength tensor
and $hgg$ interactions are induced by the top-quark loop.

If the
mass of $h$ Higgs boson is low enough for the decay
$Z_{2}\rightarrow Z_{1} h\rightarrow\bar{l}l h$ so as to be kinematically allowed, the
Tevatron bounds on this transition could severely
constrain the structure of $h$-couplings.
The relative differential distribution of the decay width
$\Gamma (Z_{2}\rightarrow\bar{l}l h)$ over the dimensionless
variable $x=(p_{l}+p_{\bar{l}})^2/M_{Z_{2}}^2$ ($p_{l}$ and $p_{\bar{l}}$
are the momenta of a lepton and antilepton, respectively) is given by
\begin{eqnarray}
\label{e39}
R_{\bar{l}lh}(x)\equiv\frac{1}{\Gamma (Z_{2}\rightarrow\bar{l}l)}
\frac{d\,\Gamma (Z_{2}\rightarrow\bar{l}l h)}{d x}\cr
=\frac{g^2\,V^{\prime}\,V\,M_{Z_{2}}^2\,\rho_{ht}^2}{4\,\cos^{2}\theta_{W}\,
\bar{V^{\prime}}\,24\,\pi^2\,v^2\,x}\left [(1-a_{h})^2+x^2-
2\,x (1+a_{h})\right ]^{\frac{1}{2}}\cr
\times\left (1-\frac{4}{x}a_{l}\right )^{\frac{1}{2}}\,
\left (1+\frac{2}{x}a_{l}\right
)\,\frac{(1-a_{h})^2+x^2+2\,x\,(2-a_{h})}{(1-x)^2+c^2}\, ,
\end{eqnarray}
where
$V^{\prime}=0.62\,\lambda^{1/2}-\theta\,g_{v}$,
$V=g_{v}+0.62\,\lambda^{1/2}\,\theta$ [27],
$\bar{V^{\prime}}=(\sqrt{3}/2)\,(1-4\,\sin^{2}\theta_{W})^{1/2}$
[31], $a_{j}=(m_{j}/M_{Z_{2}})^{2}$ with $j= l, h$;
$c=\Gamma_{Z_{2}}/M_{Z_{2}}$.
To get $\bar{V^{\prime}}$ we have taken into account the same
normalization for the $Z_{2}$-lepton couplings as the usual one
for the $Z$-lepton interplay within the SM. In the formula (\ref{e39}),
the Drell-Yan
process $\bar{p}p\rightarrow Z_{2}\rightarrow\bar{l}l$
is used as normalization. For numerical
estimations, we used the parameters determined by electroweak
data analysis [27] with pure vector couplings of $Z_{2}$.
One would expect the amplitude for the process
 $\bar{p}p\rightarrow Z_{2}\rightarrow\bar{l}l h$ to be enhanced
by the factor $\rho_{ht}^2\,V^{\prime}\,V$ with respect to the
standard model prediction.
 The calculated results of the decay width for
$Z_{2}\rightarrow \mu^{+}\mu^{-} h$ calculations
normalized to
$Z_{2}\rightarrow \mu^{+}\mu^{-} $ are given in Fig. 8
\begin{center}
\begin{figure}[h]
\resizebox{14cm}{!}{\includegraphics{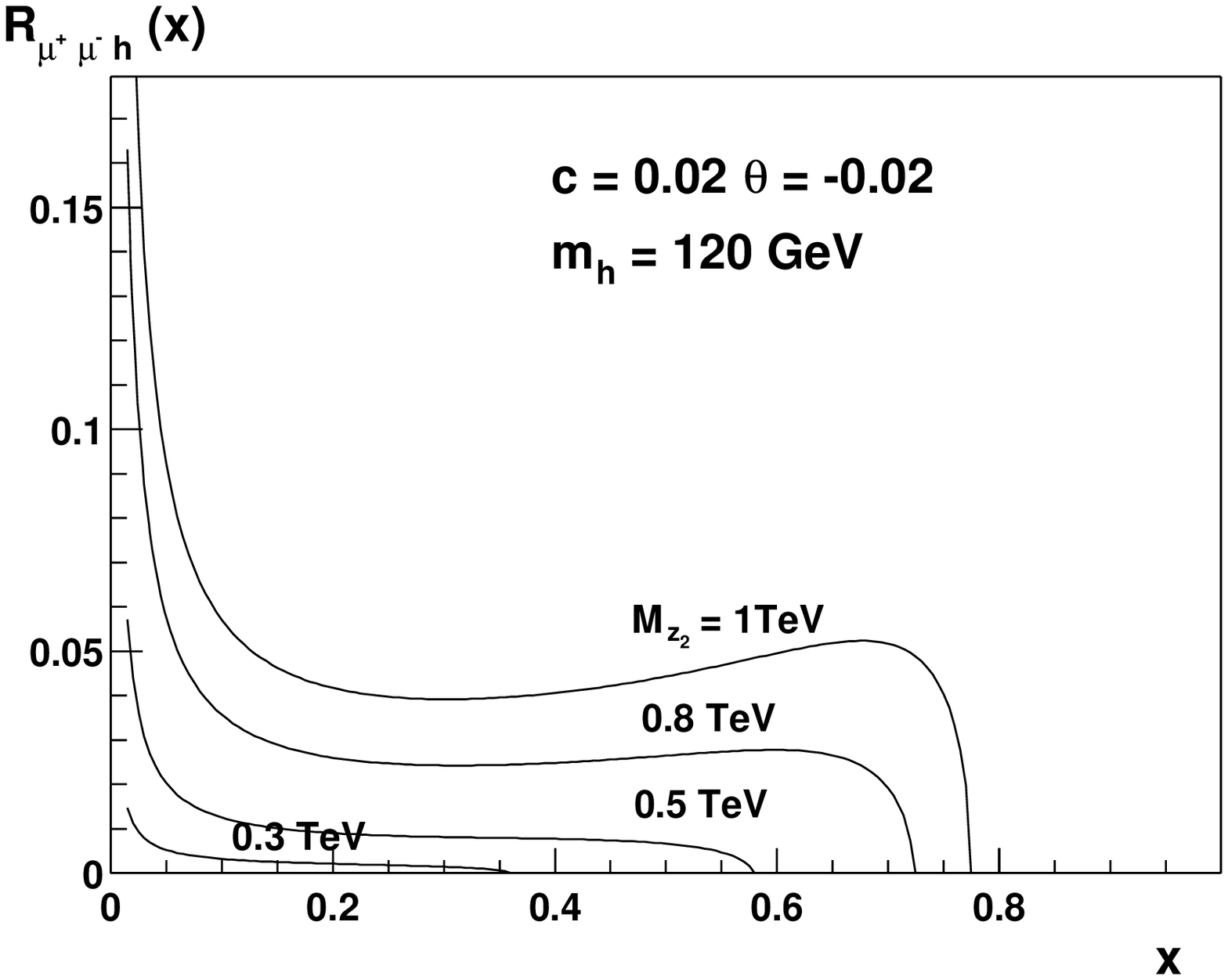}}

Fig. 8  The $x=(p_{l}+p_{\bar{l}})^2/M_{Z_{2}}^2$ distribution
of $R_{\mu^{+}\mu^{-} h}$ in the decay $Z_{2}\rightarrow \mu^{+}\mu^{-} h$
for $M_{Z_{2}}$= 0.3 TeV, 0.5 TeV, 0.8 TeV and 1 TeV.
\end{figure}
\end{center}
We used the typical value for the mixing angle $\theta$=-0.02 for
a reasonable approximation $\sin\theta\simeq\theta$ and the trial
value $c$=0.02 for the ratio $\Gamma_{Z_{2}}/M_{Z_{2}}$ [27].
Note that the changing of the parameters $\theta$ and $c$ in the
window of allowed electroweak parameters [27],
 does not lead to the visible new effects with respect to the
 estimation done in Fig. 8.

\section{Conclusion}
 To summarize, we have demonstrated that the
bounds on the scale of quark-lepton compositeness
and the $Z^{\prime}$ boson mass in the extended $SU(2)_{h}\times SU(2)_{l}$
model which are derived from the data taken at the Tevatron (CDF analysis)
can be combined to constrain the upper
limit of the masses of mass-eigenstates $\tilde{t_{1}}$ and
 $\tilde{t_{2}}$ and thus can be used to sensitively probe radiative corrections
to the MSSM Higgs sector. Comparison of experimentally
measured radiative corrections combined into $\Delta$ with its
calculations can give a precise estimation of the lower bound
on the $h$ - (as well as $A$-) Higgs boson masses.
The analysis of the scale
$\Lambda$  as well as the precise measurement of the lower bound
on the $Z^{\prime}$ boson mass at the Tevatron Run II can probe
the CP-violating mixing between two heavy neutral
CP-eigenstates $H$ and $A$,
and as a consequence, the nonminimality of the MSSM Higgs
sector. It is expected that the Tevatron Run II experiments will be able
to exclude $Z^{\prime}$ bosons with masses up to 750 GeV. This
leads to the restriction of the model scale parameter like $x=u^2/v^2$
which would grow.
Note that recent experimental limits on $W^{\prime}_{LR}$ and
 $Z^{\prime}_{LR}$ gauge bosons in the canonical left-right
symmetric model [32] require that their mass be higher than
about 800 GeV.
 An important question is whether the
forthcoming data at the Tevatron Run II at $\sqrt s$=2 TeV will
progress far enough to determine the lower bounds on the
$\Lambda$-scale and the $Z^{\prime}$ boson mass within
the models considered in this work.\\
A large decay width $h\rightarrow gg$ arises due to a sum of
one-loop intermediate heavy quark states largely decaying  into
two gluons (a new interaction of the $h$-Higgs boson is extended only
on the third family). The $h\rightarrow \bar{b}b $ channel
will
be diminished, otherwise it must have already been established
at the Tevatron. The $\bar{l}l$ mode via a gluon fusion may be
significantly enhanced in MSSM. Thus, this mode is a very
promising channel to discover CP-odd Higgs boson $A$.
Discovery region for the $\tau^{+}\tau^{-}$ mode might be enlarged
as well as
the $\mu^{+}\mu^{-}$ channel, if  a precise
reconstruction of the mass of the $A$ -Higgs bosons is available. The detection modes
$A\rightarrow\gamma\gamma h$ and $A\rightarrow ggh$
may be good channels for discovery of two Higgs bosons predicted
in the MSSM.\\
From the above consideration, we note that the Tevatron Run II
can catch an  evidence of Higgs bosons ($h$/$H$ and/or $A$) whose dominant
 final states could be
gluon jets and/or pairs of  $\tau^{+}\tau^{-}$ or
$\mu^{+}\mu^{-}$ instead of $\bar{b}b$.

We thank G. Arabidze for help in numerical calculations.

\end{document}